\documentclass[sigconf]{acmart}

\AtBeginDocument{%
  }

\usepackage{booktabs}
\usepackage{amsmath}
\usepackage{enumitem}
\usepackage{balance}

\newcommand{\Zzero}{\ensuremath{\mathrm{Z0}}}
\newcommand{\Zone}{\ensuremath{\mathrm{Z1}}}
\newcommand{\Ztwo}{\ensuremath{\mathrm{Z2}}}
\newcommand{\Zthreeauto}{\ensuremath{\mathrm{Z3}_{\mathrm{auto}}}}
\newcommand{\Zthreeval}{\ensuremath{\mathrm{Z3}_{\mathrm{val}}}}
\newcommand{\ZthreeautoSpecific}{\ensuremath{\mathrm{Z3}_{\mathrm{auto\text{-}specific}}}}
\newcommand{\Gzone}{\ensuremath{\mathrm{G}}}
\newcommand{\Uzone}{\ensuremath{\mathrm{U}}}

\copyrightyear{2026}
\acmYear{2026}
\setcopyright{cc}
\setcctype{by}
\acmConference[CIKM '26]{Proceedings of the 35th ACM International Conference on Information and Knowledge Management}{November 07--11, 2026}{Rome, Italy}
\acmBooktitle{Proceedings of the 35th ACM International Conference on Information and Knowledge Management (CIKM '26), November 07--11, 2026, Rome, Italy}
\acmDOI{10.1145/3799682.3840922}
\acmISBN{979-8-4007-2539-5/2026/11}

\begin{document}

\title{The ``Curse of Knowledge'' in LLM Query Simulation: Concept Provenance for Tracing Answer-Side Intrusion}

\author{Chenglong Ma}
\email{chenglong.ma@rmit.edu.au}
\orcid{0000-0002-6745-4029}
\affiliation{%
  \institution{RMIT University}
  \city{Melbourne}
  \state{Victoria}
  \country{Australia}
}

\author{Xinye Wanyan}
\email{xinye.wanyan@student.rmit.edu.au}
\orcid{0009-0002-7264-1803}
\affiliation{%
  \institution{RMIT University}
  \city{Melbourne}
  \state{VIC}
  \country{Australia}
}

\author{Danula Hettiachchi}
\email{danula.hettiachchi@rmit.edu.au}
\orcid{0000-0003-3875-5727}
\affiliation{%
  \institution{RMIT University}
  \city{Melbourne}
  \state{VIC}
  \country{Australia}
}

\author{Ziqi Xu}
\email{ziqi.xu@rmit.edu.au}
\orcid{0000-0003-1748-5801}
\affiliation{%
  \institution{RMIT University}
  \city{Melbourne}
  \state{VIC}
  \country{Australia}
}

\author{Jeffrey Chan}
\email{jeffrey.chan@rmit.edu.au}
\orcid{0000-0002-7865-072X}
\affiliation{%
  \institution{RMIT University}
  \city{Melbourne}
  \state{VIC}
  \country{Australia}
}

\renewcommand{\shortauthors}{Chenglong Ma, Xinye Wanyan, Danula Hettiachchi, Ziqi Xu, and Jeffrey Chan}

\begin{abstract}
LLM-generated search queries are widely used to augment IR evaluation, yet
they may contain concepts that presuppose answer-side document knowledge,
violating the information-access boundary of pre-search users.
Existing validation metrics---overlap, diversity, effectiveness---cannot
distinguish rare human-tail variation from candidate answer-side intrusion. We
introduce \emph{concept provenance}, a framework that assigns query concepts to
backstory-supported, human-central, human-tail, and candidate answer-side zones,
operationalizing a boundary that retrieval metrics alone cannot detect. Applying
concept provenance to 77,004 queries across 100 UQV100 topics, 8 LLMs, and 5
prompt conditions with two extraction pipelines (cross-pipeline token-HCIR
Spearman $\rho = 1.0$ over five condition means), we find candidate answer-side concepts constitute 7.40\% of
non-generic concepts, appearing in 97 of 100 topics with topic explaining
${\approx}67\%$ of variance. Human validation yields 68.2\% relaxed precision,
revealing a dual mechanism: knowledge intrusion (45.5\%) and deployment intrusion
(45.0\%). Diagnostic probes show these concepts carry disproportionate localized
retrieval effects (deletion $d = -0.47$, exceeding random deletion
$d = -0.34$) but explain less than 2\% of aggregate evaluation variance,
positioning concept provenance as a boundary-compliance diagnostic rather than an
evaluation-shift predictor. Under the tested conditions, no prompt condition
eliminates intrusion; post-generation concept-provenance selection achieves 99\%
elimination. \footnote{Code is available at \url{https://github.com/ChenglongMa/kcqs}}
\end{abstract}

\ccsdesc[500]{Information systems~Information retrieval}
\ccsdesc[300]{Information systems~Data management systems}
\ccsdesc[300]{Computing methodologies~Artificial intelligence}

\keywords{concept provenance, IR evaluation, diagnostic intervention, boundary compliance, information and knowledge management}

\maketitle

\vspace{-1.25em}
\section{Introduction}
\label{sec:introduction}

Information retrieval evaluation increasingly relies on query variation to assess system robustness and estimate real-world performance~\citep{bailey2015pooled,bailey2016,li2025matching}. Large language models offer a scalable approach to generating diverse query variants~\citep{alaofi2023,zendel2025}, with applications ranging from test collection building to user simulation~\citep{balog2024}. However, initial query formulation operates within a strict \emph{information-access boundary}: a user possesses only the task backstory and prior knowledge, not the search results. If an LLM-generated query contains concepts that presuppose knowledge of candidate answers, the query violates this boundary. A boundary-violating query may retrieve highly relevant documents, yet it fails as a \emph{simulation}: it does not reflect the information state of the user it purports to model.

Consider UQV100~\citep{bailey2016} topic 036 (mad cow disease symptoms in humans). The backstory describes a friend worried about health after a UK trip during the outbreak. A human with this context might query ``mad cow disease symptoms in humans''—all concepts trace to the backstory. An LLM instead produces ``NHS guidance prion disease symptoms classic vs variant Creutzfeldt-Jakob''—deploying biomedical terms absent from all observed human queries and the backstory. These are \emph{candidate answer-side concepts} (\Zthreeauto{}): information the user is \emph{trying to find}, not information available before searching. Meanwhile, human queries occasionally include rare terms like ``affect'' (prevalence $< 2\%$)—natural vocabulary variation.

Existing query validation approaches based on lexical overlap, diversity metrics, or retrieval effectiveness~\citep{alaofi2023,zendel2025,ran2025,breuer2024,kruff2026} cannot distinguish human-tail variation from answer-side intrusion. Both produce unusual, high-IDF terms; both diverge from typical human queries; both may improve or degrade retrieval depending on the system. Yet one reflects plausible user behavior, while the other reflects a boundary violation. Without this distinction, synthetic IR evaluation~\citep{rahmani2024,rahmani2025,balog2025} cannot diagnose whether LLM queries that shift retrieval pools, alter judged coverage, and produce divergent evaluation outcomes do so because of answer-side intrusion or because of legitimate vocabulary variation—even when aggregate retrieval metrics appear reasonable.

We introduce \emph{concept provenance}, a framework that assigns query concepts to provenance zones—backstory-supported (\Zzero{}), human-central (\Zone{}), human-tail (\Ztwo{}), and candidate answer-side (\Zthreeauto{}) through a priority-ordered protocol with two-stage validation (\Zthreeauto{} automatic, \Zthreeval{} manual). We operationalize the framework on UQV100 with 77,004 queries from 8 LLMs under 5 prompt conditions, assessing construct validity through dual pipelines, threshold sensitivity, and human annotation.

This paper contributes: (1)~a concept-provenance framework decomposing query concepts into backstory-supported, human-central, human-tail, and candidate answer-side zones under an explicit information-access boundary; (2)~an empirical \Ztwo{}/\Zthreeauto{} distinction validated through dual pipelines, manual annotation, and sensitivity analysis; (3)~diagnostic probes showing \Zthreeauto{} concepts carry disproportionate localized retrieval signal ($R^{2} < 2$\% at aggregate level)—establishing concept provenance as a boundary-violation diagnostic; and (4)~a boundary compliance analysis showing prompt-based mitigation reduces but does not eliminate intrusion, while post-generation selection can nearly eliminate detected intrusions under the tested candidate pool.

\noindent Three research questions organize the investigation:
\begin{enumerate}[label=\textbf{RQ\arabic*:},leftmargin=*,nosep]
\item How are concepts in LLM-generated initial queries distributed across provenance zones, and how reliably can the zones be operationalized?
\item Does candidate answer-side intrusion predict shifts in retrieval pools, judged coverage, and system rankings relative to the observed human-query distribution?
\item Does prompt-based mitigation suffice to keep LLM queries within human-range boundary compliance, or is post-generation validation required?
\end{enumerate}

\section{Related Work}
\label{sec:related-work}

\subsection{Human and LLM-Generated Query Variants}
\label{sec:rw-query-variation}

Human query variability reflects heterogeneous behavior and directly shapes IR
evaluation outcomes~\citep{abuonq2026heterogeneous,bailey2015pooled,bailey2016}.
Work on UQV100 and pooled evaluation demonstrates that
query variation among human searchers affects system
rankings as strongly as system variation, with different workers formulating
queries that retrieve substantially different document sets for the same
information need. The observed human-query
distribution therefore serves as a reference against which synthetic query generators must be
validated.

LLMs now serve as a practical source of query variants for evaluation.
Alaofi et al.~\citep{alaofi2023} show that generative LLMs can
produce diverse query variants from information need descriptions, reporting
substantial pool overlap with human queries alongside systematic
differences in linguistic features and retrieval behavior.
Zendel et al.~\citep{zendel2025} extend this analysis to a comparative study
of linguistic and retrieval diversity, finding that LLM-generated queries
exhibit higher syntactic variability than human queries but produce different
evaluation outcomes, and concluding that LLMs are not yet a suitable
replacement for human searchers in IR evaluation research.
Further work on data fusion of LLM variants with human
queries~\citep{ran2025,breuer2024} confirms that LLM queries differ from
human queries in ways that affect evaluation.

These studies validate LLM query variants along dimensions of diversity, pool
overlap, and retrieval effectiveness---but treat all concept-level differences
between LLM and human queries uniformly. A concept that is rare among humans
but attested~(\Ztwo{}, human-tail) and a concept entirely absent from observed
human variants yet salient in relevant documents~(\Zthreeauto{}, candidate
answer-side) are indistinguishable under existing overlap and diversity
metrics. This paper asks a different question: not whether LLM queries are
diverse or effective, but where their concepts originate relative to the
backstory and the observed human initial-query distribution.

\subsection{Synthetic Evaluation Bias and LLM Evaluation Artifacts}
\label{sec:rw-synthetic-bias}

Several studies document systematic biases that LLM-generated
artifacts introduce into IR evaluation~\citep{rahmani2024,rahmani2025,balog2025,dai2024neural,li2025preference}.
Rahmani et al.~\citep{rahmani2024,rahmani2025} demonstrate that synthetic queries, documents, and relevance
labels can produce evaluation outcomes that diverge from
human-based benchmarks, with system ranking correlation as a primary concern;
the bias spans multiple evaluation components.
Balog et al.~\citep{balog2025}, Dai et al.~\citep{dai2024neural}, and Li et al.~\citep{li2025preference} confirm at the ecosystem level that biases propagate across
evaluation components---rankers, retrievers, judges, and labels---making it urgent to
understand where and how LLM-generated content distorts evaluation.

These results connect to a broader distributional-alignment problem:
synthetic artifacts may be useful or fluent while still occupying a
different behavioral distribution from the human artifacts they
replace~\citep{ahmed2026comparative}.
For initial-query simulation, alignment is not only lexical or metric-based;
the generated query must also respect the simulated user's information state.
Leakage and contamination work in LLM evaluation similarly warns that
downstream components can inherit information unavailable in the intended
evaluation setting~\citep{balog2025,li2025preference}. We study the query-side
version of that problem: answer-side concepts entering the query before any
retrieval interaction has occurred.

These results show that synthetic evaluation artifacts can shift
system rankings and effectiveness estimates, but they do not decompose the
query-side mechanisms responsible. This paper isolates one specific
mechanism: candidate answer-side concept intrusion, where LLM-generated
initial queries contain document-salient concepts absent from observed
human initial-query variants and unsupported by the backstory. Diagnostic
deletion and injection probes with matched controls provide evidence for
this mechanism beyond query length and specificity confounds.

\subsection{Query Simulation Validation and Knowledge-Conditioned Generation}
\label{sec:rw-simulation-validation}

Prior work surveys user simulation validation along axes of behavioral
realism, statistical plausibility, and evaluation utility~\citep{balog2024},
and proposes taxonomies and auditable frameworks for query and user
simulation~\citep{kruff2026,ma2026verifiable}.
Alaofi et al.~\citep{alaofi2022} discuss query
formulation sources conceptually but do not operationalize provenance-based
validation. Knowledge- and profile-conditioned
generation~\citep{alaofi2025,ma2025pub,wanyan2025temporal,wanyan2026taskaware}---conditioning
simulated behavior on demographic, personality, temporal, or task-aware
attributes---motivates
knowledge-level prompts as experimental conditions but does not validate
whether generated concepts respect the information-access boundary.

Concept provenance provides a complementary axis: existing validation
measures assess whether generated queries \emph{resemble} human queries but
do not ask whether each concept \emph{could plausibly have been formulated}
given only the backstory. The distinction between human-tail
variation~(\Ztwo{}) and candidate answer-side intrusion~(\Zthreeauto{}) is
invisible to similarity, diversity, and effectiveness metrics.

\begin{figure*}[t]
  \centering
  \includegraphics[width=0.8\linewidth]{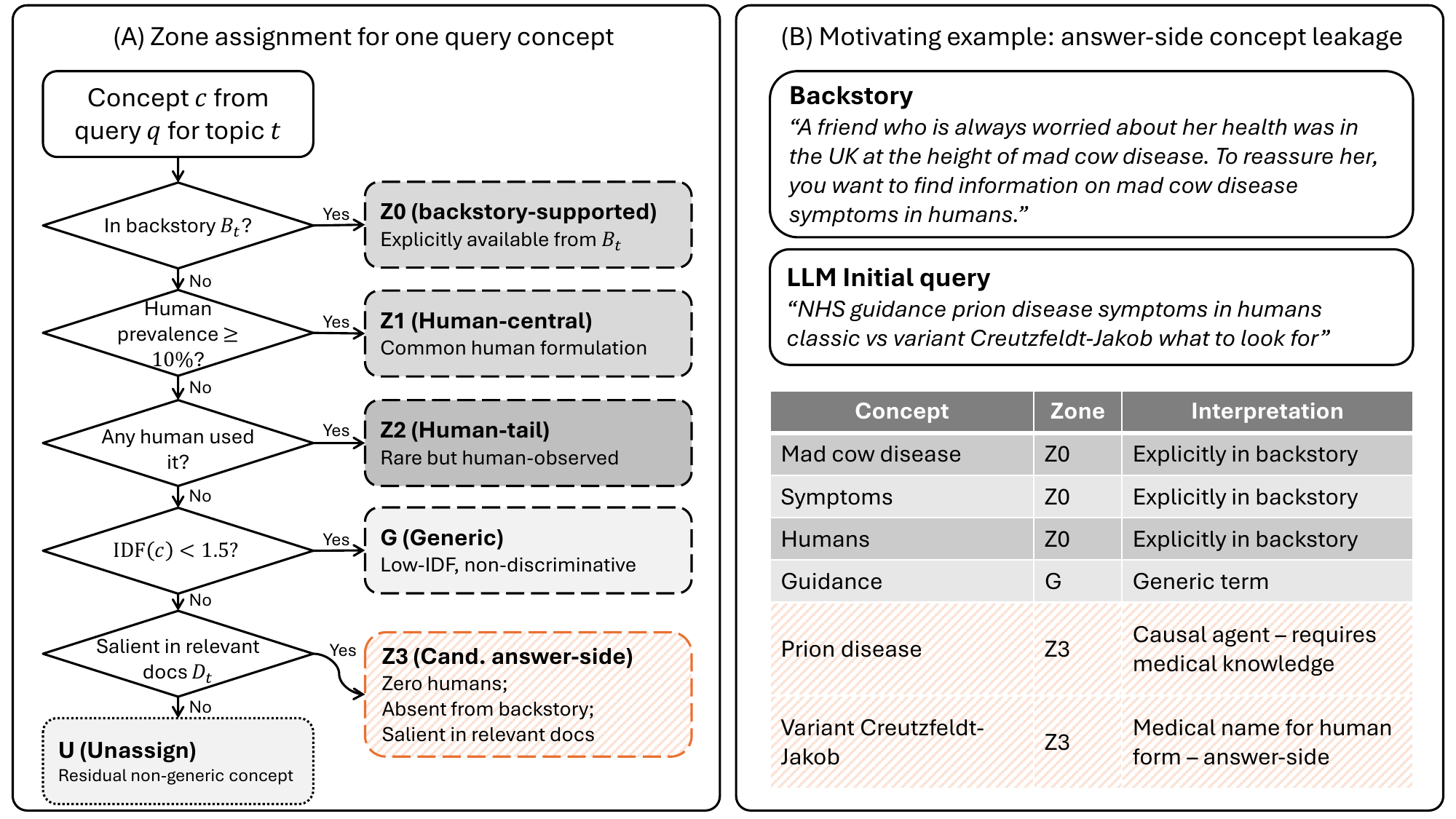}
  \caption{
    Concept-Provenance Framework for assigning each concept in an LLM-generated initial query to a topic-specific provenance zone.
    Panel~(A) shows the priority-ordered assignment procedure: backstory-supported concepts are assigned first, followed by human-central concepts, human-tail concepts, generic concepts, candidate answer-side concepts, and residual unassigned concepts.
    The diagram abbreviates the multi-signal generic decision as an IDF test; Table~\ref{tab:zones} and \S\ref{sec:extraction} specify the complete filter used to assign \Gzone{}, including the ``guidance'' example in Panel~(B).
    Panel~(B) illustrates the framework with a mad cow disease-search scenario, where the user initially knows only the cow disease and symptoms.
    The example highlights the distinction between human-attested rare \Ztwo{} concepts and \Zthreeauto{} concepts that are absent from the backstory and human initial queries but salient in relevant documents, resembling terms a user might acquire only after inspecting search results and reformulating.
  }
  \Description{
    A two-panel figure. The left panel shows a top-to-bottom flowchart for assigning a query concept to \Zzero{}, \Zone{}, \Ztwo{}, \Gzone{}, \Zthreeauto{}, or \Uzone{}. The right panel shows a mad cow disease-search example where disease-related terms such as BSE and prion are highlighted as candidate answer-side concepts.
  }
  
  \label{fig:flowchart}
\end{figure*}

\section{Concept-Provenance Framework}
\label{sec:framework}

Initial query formulation in information retrieval operates within a strict
information-access boundary~\citep{azzopardi2007,balog2024}: the searcher
possesses only the backstory~$B_t$ describing their information need, with no exposure to search
results or documents. This boundary defines what concepts a pre-search user
could plausibly formulate. For query simulation to support valid IR evaluation,
generated queries must respect this boundary. When a query contains concepts
that require document-side knowledge---concepts absent from the backstory and
from the observed human-query distribution, yet salient in relevant
documents---the simulated searcher has accessed information outside the
boundary. Concept provenance operationalizes this distinction, providing a validation axis for
LLM-generated initial search queries.

\subsection{Concepts and Provenance Zones}
\label{sec:zones}

A \emph{concept} is a normalized semantic unit extracted from queries,
backstories, or documents. Candidate units include named entities, noun
phrases, domain-specific terms, lemmatized content words, and informative
n-grams, following prior IR work that models and weights query concepts
as retrieval units~\citep{bendersky2008discovering,metzler2007latent}.
For a topic~$t$, we define $B_t$~as the backstory (information need
description), $H_t$~as the set of observed human initial-query
variants~\citep{bailey2016}, $D_t$~as the set of judged relevant documents, and
$Q_t^{\text{LLM}}$~as the set of LLM-generated initial queries under various
conditions. The extraction pipeline and normalization protocol are detailed
in~\S\ref{sec:methodology}; the framework specifies only that concepts are
comparable units across all sources.

Concepts are assigned to provenance zones based on their relationship to
the information-access boundary. Table~\ref{tab:zones} defines the zones
with operational criteria. Zone assignment follows a strict priority order:
\(\Zzero > \Zone > \Ztwo > \Gzone > \Zthreeauto\). A concept satisfying multiple criteria
receives the highest-priority label. Evidence-based labels (backstory
support, human attestation) take precedence over the generic
heuristic---observed evidence is authoritative over statistical frequency.
The generic filter~(\Gzone{}) guards only the~\Zthreeauto{} path, using corpus
IDF and general-language frequency~(\S\ref{sec:methodology}) to prevent
non-discriminative terms from reaching~\Zthreeauto{}. The residual
category~\Uzone{} captures non-generic concepts matching no zone; these are
included in the HCIR denominator to avoid inflating intrusion.

\begin{table}[t]
\centering
\caption{Concept-provenance zone definitions. All labels are topic-specific
  and assigned with priority \(\Zzero > \Zone > \Ztwo > \Gzone > \Zthreeauto\). Worker-level
  prevalence is the proportion of workers who used the concept for topic~$t$.
  Central threshold default: 10\%. Strict absence default: worker count~=~0.}
\label{tab:zones}
\footnotesize
\setlength{\tabcolsep}{4pt}
\begin{tabular}{@{}p{0.08\linewidth}p{0.22\linewidth}p{0.62\linewidth}@{}}
\toprule
\textbf{Zone} & \textbf{Label} & \textbf{Operational Definition} \\
\midrule
\Gzone{} &
  Generic &
  Matches a stopword or curated task-generic term, has corpus IDF below 1.5,
  or is a non-entity unigram with wordfreq Zipf frequency at least 4.5.
  Applied only to concepts with no backstory or human evidence;
  guards the~\Zthreeauto{} path. \\
\addlinespace
\Zzero{} &
  Backstory-supported &
  Explicitly appears in or is a close paraphrase of~$B_t$.
  Conservative: explicit lexical presence only, not inferred entailment. \\
\addlinespace
\Zone{} &
  Human-central &
  Worker-level prevalence in $H_t \geq$ central threshold (default 10\%). \\
\addlinespace
\Ztwo{} &
  Human-tail &
  Non-zero worker-level prevalence in~$H_t$, below central threshold.
  Within the observed range of human initial-query formulation. \\
\addlinespace
\Zthreeauto{} &
  Candidate answer-side &
  Must satisfy \textbf{all five} conditions:
  (a)~absent in~$H_t$ (worker count~=~0),
  (b)~not backstory-supported,
  (c)~salient in relevant documents (top-200 TF-IDF, $\geq 2$ docs),
  (d)~not in the stopword or curated task-generic lists and IDF~$\geq 1.5$,
  (e)~not contextually generic by the general-language frequency
  filter~(\S\ref{sec:methodology}). \\
\addlinespace
\Uzone{} &
  Unassigned &
  Non-generic, non-backstory, non-human-attested, non-document-salient.
  Residual category; included in HCIR denominator. \\
\addlinespace
\Zthreeval{} &
  Validated &
  Manually confirmed subset of~\Zthreeauto{} via human annotation. \\
\bottomrule
\end{tabular}
\end{table}

The~\Zthreeauto{} label is assigned by automatic rules and carries a known
false-positive risk: extraction errors, paraphrases missed by matching, or
backstory-supported concepts not recognized as such may be mislabeled. We
therefore employ a two-stage protocol. All primary analyses use~\Zthreeauto{}
for scalable detection. A subset of~\Zthreeauto{} concepts is then subjected
to manual validation, producing~\Zthreeval{}, the manually confirmed
candidate answer-side subset. Key analyses are replicated on~\Zthreeval{} to
assess conclusion stability under precision-corrected conditions.
The~\Zthreeauto{} precision, inter-annotator agreement, and error taxonomy
are reported as construct-validity evidence~(\S\ref{sec:exp1-concept-provenance}).

\subsection{Intrusion Measurement}
\label{sec:hcir}

We define the \emph{Hidden Concept Intrusion Rate} (HCIR) as the proportion
of a query's non-generic concepts that fall in the candidate answer-side zone:
\begin{equation}
\label{eq:hcir}
\text{HCIR}(q, t)
  = \frac{|\text{Concepts}(q) \cap \Zthreeauto(t)|}
         {|\{c \in \text{Concepts}(q) : c \notin \Gzone(t)\}|}.
\end{equation}
The denominator counts all non-generic concepts in query~$q$, including
unassigned~(\Uzone{}) concepts, to prevent denominator manipulation.
We also define \emph{HCI-Presence}$(q,t)$ as a binary indicator of
any~\Zthreeauto{} concept being present, the \emph{Human-Tail Rate}
(HTR) with~\Ztwo{} in the numerator, and the \emph{Human-Central Rate}
(HCR) as the proportion of non-generic concepts in~\Zone{}. A token-level variant decomposes
concepts into tokens, neutralizing granularity differences across
pipelines; within-pipeline analyses use concept-level HCIR as the
pre-registered primary metric, cross-pipeline comparisons use token-level.
Concept-level HCIR is sensitive to denominator dilution from query
length; HCI-Presence and token-level HCIR serve as length-robust
complements.

\textit{Prevalence and thresholds.}
Worker-level prevalence---the proportion of workers who used a concept for
topic~$t$---is the primary measure of human concept usage, robust to
within-worker repetition. The central threshold is 10\% prevalence,
separating human-central~(\Zone{}) from human-tail~(\Ztwo{}). The
strict~\Zthreeauto{} absence criterion requires worker count~=~0; a
sensitivity variant relaxes this to near-absent ($\leq 1$). Defaults are
pre-registered; sensitivity
analysis~(\S\ref{sec:exp1-concept-provenance}) varies them systematically.

\textit{Document salience.}
Candidate answer-side concepts must be salient in relevant documents;
operational criteria and sensitivity analysis are detailed
in~\S\ref{sec:salience}.

\textit{Cross-source matching.}
Concepts extracted from queries, backstories, and documents are matched
via lemmatized exact matching, which minimizes
false matches at the cost of recall. Sensitivity checks include fuzzy string matching and
embedding-based similarity; full matching rules are specified
in~\S\ref{sec:methodology}.

Because human absence is part of the~\Zthreeauto{} definition, this paper's
empirical claims focus on non-circular axes: absolute rates, condition and
topic variation, and retrieval effects of~\Zthreeauto{} in LLM
queries---not human-versus-LLM superiority. Circularity implications are
discussed in~\S\ref{sec:discussion}.

\section{Methodology}
\label{sec:methodology}

We operationalize the concept-provenance framework
(\S\ref{sec:framework}) for reproducible analysis on a multi-variant
query test collection. All thresholds were
pre-registered before full-scale experiments; sensitivity analysis
(\S\ref{sec:exp1-concept-provenance}) varies each parameter
systematically.

\subsection{Data and Search Tasks}
\label{sec:data}

We used UQV100~\citep{bailey2016}, a test collection of 100 topics
drawn from the TREC Web Track 2009--2014. Each topic includes a
backstory describing the information need and approximately
50~crowd-sourced initial search queries. Workers saw only the backstory
before formulating queries, with no exposure to search results or
relevant documents---an elicitation protocol that
operationalizes the information access boundary
(\S\ref{sec:framework}).
In total, 263~crowd workers contributed 10{,}835 query variants
(5{,}744 unique after normalization), averaging 108~per topic.
The document collection is ClueWeb12-B13, with 55{,}587 graded
relevance judgments (scale 0--4) across 54{,}880 unique documents
(median 98.5 relevant documents per topic, mean 114.8).

\subsection{Query Generation}
\label{sec:generation}

LLM queries were generated under five prompt conditions. Three used
only the backstory: \emph{generic} (no knowledge instruction),
\emph{high-knowledge} (domain-expert framing), and
\emph{boundary-constrained} (explicit instruction and few-shot example
to avoid unsupported concepts). Two oracle controls intentionally
exposed document-side evidence: \emph{oracle-text} added TF-IDF-selected~\citep{salton1988term}
relevant-document excerpts, and \emph{oracle-terms} added the top-10
TF-IDF terms from relevant documents. These oracle conditions are
positive controls for document-aware language, not simulations of
pre-search users.

The generation set covers eight LLMs from OpenAI, Anthropic, and
DeepSeek, spanning GPT-4.1/GPT-5.4 families, Claude Sonnet, and
DeepSeek-V3.1. All models used temperature~$T{=}1.0$,
$\text{top\_p}{=}1.0$, and maximum 500~tokens; 20~candidate queries
were generated for each topic--condition--model cell with up to three
deduplication retries, yielding 77{,}004 analyzed LLM queries. Exact
model identifiers, prompt templates, generation dates, and retry logs
are provided in the shared codebase.

\subsection{Concept Extraction}
\label{sec:extraction}

Concepts were extracted through two complementary pipelines to
support cross-pipeline robustness
analysis~(\S\ref{sec:exp1-concept-provenance}). Both pipelines
processed queries, backstories, and relevant documents under a
shared normalization protocol (lowercase, lemmatization, punctuation
removal).

\textit{Pipeline~A (lexical/statistical).}
Content words were lemmatized using spaCy~\citep{honnibal2020spacy}
(\texttt{en\_core\_web\_sm}). Informative bigrams were identified via pointwise mutual
information (PMI) filtering~\citep{church1990word}. TF-IDF
terms from queries and documents supplemented the unigram inventory.

\textit{Pipeline~B (entity/phrase-based).}
Named entities and noun phrases were extracted via spaCy's NER and
dependency parser~\citep{honnibal2020spacy}. Domain-specific multi-word terms were retained
with a minimum two-character filter.

\textit{Generic filtering.}
Both pipelines share a multi-signal generic filter. A concept is
classified as generic and excluded from the~\Zthreeauto{} path if it
meets any of: (1)~membership in a standard stopword or curated
task-generic term list (${\sim}170$~terms); (2)~corpus
IDF~$< 1.5$; or (3)~high general-language frequency
(wordfreq Zipf frequency~$\geq 4.5$; \citep{speer2022wordfreq}) for single-token concepts that are not
named entities---multi-token phrases are exempt from criterion~(3).
Evidence-based labels (\Zzero{}, \Zone{}, \Ztwo{}) override the generic
heuristic: backstory-confirmed or human-attested concepts retain
their provenance label regardless of corpus
frequency~(\S\ref{sec:zones}).

Cross-pipeline comparisons use token-level HCIR to neutralize the
granularity difference (${\sim}1.3$ vs.\ ${\sim}2.1$~tokens per
concept); within-pipeline analyses use concept-level
HCIR~(\S\ref{sec:exp1-concept-provenance}).

\subsection{Zone Assignment}
\label{sec:zone-assignment}

Labels follow the priority order in~\S\ref{sec:zones}. Pre-registered
defaults: central threshold 10\% worker prevalence, strict
\Zthreeauto{} absence (count~$= 0$), document salience top-200
TF-IDF ($\geq 2$ documents), generic IDF\,$< 1.5$, lemmatized exact
matching. Sensitivity variants: central $\in \{5, 15, 20\}\%$,
near-absent ($\leq 1$), $k \in \{10, 20, 50, 100\}$, IDF
$\in \{1.0, 2.0\}$, fuzzy matching.

\subsection{Document Salience}
\label{sec:salience}

A concept was classified as document-salient if it ranked within the
top-$k$ TF-IDF terms~\citep{salton1988term} over~$D_t$ and appeared in $\geq 2$ relevant
documents. The primary threshold was $k = 200$; sensitivity analysis
varied $k$ across $\{10, 20, 50, 100\}$~(\S\ref{sec:exp1-concept-provenance}).
Document-salient concepts were classified into IDF quality tiers:
\emph{specific} (IDF~$\geq 3.0$), \emph{moderate}
($1.5 \leq \text{IDF} < 3.0$), and \emph{generic}
(IDF~$< 1.5$, excluded before reaching \Zthreeauto{}).
Because $D_t$ comprises only judged relevant documents, this
operationalization is conservative---biasing toward under-detection.
A retrieved-document salience variant is specified in the shared codebase.

\subsection{Human Validation Protocol}
\label{sec:validation}

A stratified sample of 400 query--concept pairs (220~\Zthreeauto{},
180~other zones) from 92~topics was annotated by three human annotators.
Each annotator classified concept origin as \emph{backstory},
\emph{common\_know\\ledge}, \emph{personal\_experience}, or
\emph{requires\_research}, seeing only backstory, query, and concept,
calibrated against a general pre-search user. Majority vote with
adjudication produced consolidated labels. We report strict precision
(\emph{requires\_research}) and relaxed precision
(\emph{personal\_experience}+\emph{requires\_research}). Annotation
guidelines and calibration examples are in the shared codebase.

\subsection{Retrieval and Evaluation Setup}
\label{sec:retrieval}

Core retrieval runs use three systems: BM25~\citep{robertson2009probabilistic} ($k_1{=}0.9$, $b{=}0.4$) as
primary, BM25+RM3~\citep{lavrenko2001relevance}, and QL~\citep{ponte1998language} ($\mu{=}2500$).
As a neural scoring robustness check, we additionally report a cross-encoder
reranker~\citep{nogueira2019passage} (ms-marco-MiniLM-L-12-v2) re-scoring
BM25 top-100 candidates.
Metrics include nDCG@10~\citep{jarvelin2002cumulated},
Recall@1K, bpref~\citep{buckley2004bpref}, and
RBP~\citep{moffat2008rank}. Given incomplete UQV100
qrels~\citep{bailey2016}, primary outcomes emphasize pool
overlap, judged ratio, and system ranking correlation
(Kendall's~$\tau$~\citep{kendall1938new}).

\subsection{Statistical Analysis}
\label{sec:statistical}

Topic-level averages are the primary unit of analysis. Pairwise
condition comparisons use paired bootstrap tests~\citep{efron1979bootstrap}; multi-factor analyses
use mixed-effects regressions~\citep{bates2015fitting} controlling for condition and query
length with topic/model effects where applicable.
Holm correction~\citep{holm1979simple} is used for multiple comparisons, and key condition
comparisons report Cohen's~$d$ effect sizes~\citep{cohen1992power} with
95\% bootstrap confidence intervals where available.
Full statistical specifications
and scripts are in the shared codebase.

\section{Experiment 1: Concept Provenance}
\label{sec:exp1-concept-provenance}

This experiment addresses RQ1 by characterizing concept-provenance
zone distributions across prompt conditions, models, and topics on
the full UQV100 collection~\citep{bailey2016}. Results are from
Pipeline~A (100~topics, five conditions, eight models) unless
otherwise noted; cross-pipeline comparisons use token-level
HCIR~(\S\ref{sec:extraction}).

\subsection{Findings}
\label{sec:exp1-findings}

\begin{figure*}[t]
  \centering
  \includegraphics[width=0.86\textwidth]{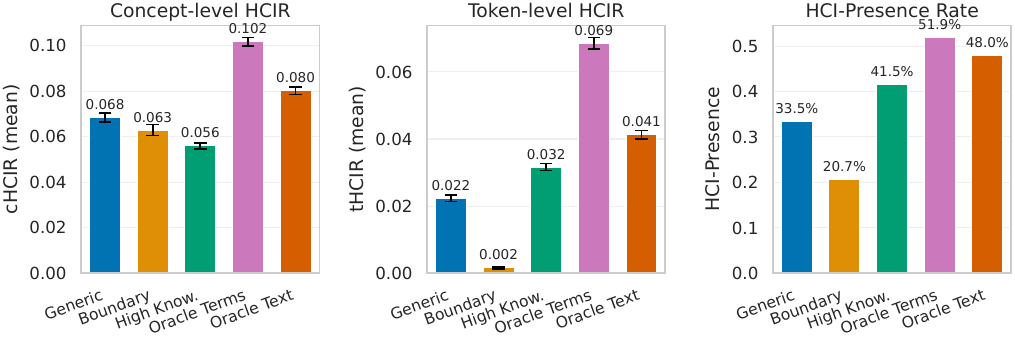}
  \caption{Condition-level intrusion profile across concept-HCIR, token-HCIR,
    and HCI-Presence. High-knowledge prompts lower concept-HCIR but increase
    token-HCIR and presence; boundary prompts reduce token-level intrusion,
    while oracle controls remain highest under presence-oriented views.}
  \Description{Three bar charts comparing generic, boundary, high-knowledge,
    oracle-terms, and oracle-text conditions. The first shows concept-level
    HCIR, the second token-level HCIR, and the third HCI-Presence rate. The
    boundary condition is low in token-HCIR and HCI-Presence, while oracle
    conditions are high.}
  \label{fig:exp1-condition-gradient}
\end{figure*}

Across all conditions and models, candidate answer-side concepts
(\Zthreeauto{}) constitute 7.40\% of non-generic query concepts
(HCIR) and occur in 97 of 100~topics (51{,}363 query--concept pairs).
The multi-signal filter~(\S\ref{sec:zone-assignment}) excludes
non-discriminative terms from the candidate answer-side path, so all
reported \Zthreeauto{} concepts are at least moderate by the IDF tiering
used for human validation.

The condition gradient is consistent across metrics
(Figure~\ref{fig:exp1-condition-gradient}): high-knowledge (5.60\%) $<$
boundary (6.28\%) $<$ generic (6.84\%) $<$ oracle-text (8.02\%)
$<$ oracle-terms (10.16\%). Relative to the generic baseline,
high-knowledge prompts yield a moderate concept-HCIR reduction
(Cohen's $d = -0.12$) and boundary a smaller one ($d = -0.04$),
while oracle-terms produces the highest \Zthreeauto{} density,
consistent with explicit document-term exposure providing more
answer-specific vocabulary. A mixed-effects model with random
topic intercepts (ICC\,=\,0.41) confirms that topic explains
approximately 41\% of concept-HCIR variance at the query level;
adding model as a fixed effect contributes
$\Delta R^{2} < 0.001$.

The generic--high-knowledge comparison reveals a metric divergence: concept-HCIR
is \emph{lower} under high-knowledge (5.60\% vs.\ 6.84\%,
$d = -0.12$, $p < 10^{-5}$), while HCI-Presence diverges in the
opposite direction (41.5\% vs.\ 33.5\%, $p < 0.001$).
High-knowledge prompts produce more queries \emph{containing}
\Zthreeauto{} concepts but at lower per-query density, a
denominator-dilution effect driven by longer queries
(17.9 vs.\ 12.2~tokens).

A pre-registered confounder regression
(HCIR $\sim$ condition $+$ length $+$ (1\,|\,topic)) confirms that
length is not a positive confounder ($\beta = -0.0005$, reflecting
dilution). Condition effects survive all controls: boundary vs.\
generic $\beta = -0.006$ ($p < 10^{-7}$), high-knowledge vs.\
generic $\beta = -0.010$ ($p < 10^{-16}$); adding model contributes
$\Delta R^{2} < 0.001$.

\textit{Model and topic variation.}
Across eight models, the fraction of queries containing zero
\Zthreeauto{} concepts ranges from 61.6\% (GPT-5.4-nano) to
73.6\% (DeepSeek-V3.1), with a pattern suggestive of a non-monotonic
relationship with model capability under the tested conditions:
GPT-5.4 produces \Zthreeauto{} concepts more frequently than
GPT-4.1 (\Zthreeauto{}-free rates of 62.6\% vs.\ 71.9\%),
consistent with more capable models drawing more heavily on
parametric knowledge during query formulation. Despite this per-query variation, a three-factor variance decomposition
(topic$\times$condition$\times$model cell means) confirms that
topic accounts for approximately 67\% of concept-HCIR variance,
while both condition ($<$\,1\%) and model ($<$\,1\%) contribute
negligibly---candidate answer-side intrusion is primarily a
property of the information need, with prompt design providing
statistically significant but small reductions
(Figure~\ref{fig:exp1-condition-gradient}). Topic-level
variation is wider (0.0\%--17.2\%; median 5.78\%, IQR [3.9\%,
7.2\%]), reflecting genuine differences in how much
document-specific vocabulary each information need elicits from LLMs.

\subsection{Robustness}
\label{sec:exp1-robustness}

\textit{Cross-pipeline validation.}
Token-level HCIR should be interpreted separately from HCI-Presence:
the former is a token fraction, while the latter is the percentage of
queries containing at least one \Zthreeauto{} concept. It resolves the
granularity difference between pipelines (Pipeline~A\,=\,0.034,
B\,=\,0.043), with perfect
condition-ranking agreement (Spearman $\rho = 1.0$ over five
means~\citep{spearman1904proof})
and both preserving the
boundary~$\ll$~generic-range~$\ll$~oracle pattern.

\textit{Human validation.}
A stratified annotation study~(\S\ref{sec:validation}) assessed
\Zthreeauto{} construct validity on 220~query--concept pairs from
92~topics. Three annotators independently classified each pair's
concept origin---seeing only the backstory, query, and
concept---calibrated against a general pre-search user with no prior
exposure to relevant documents or search results; labels were
consolidated by majority vote with discussion-based adjudication.
Strict precision (\emph{requires\_research} only) is 45.5\%, below
the pre-registered 50\% threshold (weakened tier, 30--50\%);
relaxed precision
(\emph{personal\_experience}~$+$~\emph{requires\_research}) reaches
68.2\% (moderate tier). Among the high-specificity subset
(\ZthreeautoSpecific{}, IDF\,$\geq 3.0$), relaxed precision rises
to 78.2\%, approaching the strong tier. We therefore treat
\Zthreeauto{} as a high-recall diagnostic signal rather than a
validated ground-truth label; key analyses are replicated on
\Zthreeval{} and \ZthreeautoSpecific{} to assess conclusion stability.
The gap between strict and relaxed precision reflects two complementary
intrusion mechanisms: \emph{knowledge intrusion} (45.5\%---concepts
genuinely unfamiliar to a general user) and \emph{deployment intrusion}
(45.0\%---concepts accessible but not naturally query-formulated); the
remaining 9.5\% are backstory false negatives reclassifiable via
BCC~(\S\ref{sec:exp1-robustness}). Per-condition validation preserves the
boundary~$\ll$~generic-range~$\ll$~oracle structure under restriction to
validated items. The dual-mechanism decomposition is interpreted
in~\S\ref{sec:discussion}.

\textit{Annotation bias.}
LLM annotators are not reliable substitutes for human judgment:
human--LLM agreement is only fair ($\kappa=0.21$--$0.37$), whereas
LLM--LLM agreement reaches $\kappa=0.791$, indicating a shared
knowledge-projection bias~(\S\ref{sec:discussion}).

\textit{Threshold sensitivity.}
Document salience stringency is the primary sensitivity axis: tightening
the TF-IDF cutoff from $k = 200$ to $k = 10$ reduces HCIR from 6.2\% to
1.9\%, while relaxing absence to near-absent ($\leq 1$~worker) adds
${\sim}13\%$ more \Zthreeauto{} concepts. The
boundary~$\ll$~generic-range~$\ll$~oracle structure is preserved across
all ten tested parameter combinations; cross-pipeline agreement holds under
all settings ($\rho = 1.0$ token-level over five condition means). The
defaults are operating points rather than fitted optima: 10\% worker
prevalence separates central from tail human usage; strict absence protects
the \Ztwo{}/\Zthreeauto{} boundary; and top-200 salience balances the top-50
false-negative behavior observed in pilots against the generic-IDF filter.
Because incomplete qrels bias toward under-detection, the $k$ sweep tests
salience stringency within judged relevant documents, while a retrieved- or
contrastive-document salience variant remains the appropriate future check
for source sensitivity. The stability of the condition gradient across $k$
values provides indirect evidence that qrel incompleteness does not drive
the observed condition-level patterns.

\textit{Backstory compositional coverage (BCC).}
A BCC-aware matching mode that credits concepts whose constituent tokens
all appear individually in the backstory reduces the \Zthreeauto{} pool by
10.7\% and mean HCIR by 15\% (0.0740\,$\to$\,0.0630), with the largest
reduction for boundary-prompt queries. We retain strict matching as the
primary operationalization for \Zzero{} precision; BCC confirms that
compositional backstory coverage is a tractable refinement.

Having established that candidate answer-side concepts appear at
non-trivial, condition-dependent rates and that the
\Ztwo{}/\Zthreeauto{} distinction is robust across pipelines and
thresholds, we next examine whether this intrusion is associated
with measurable shifts in retrieval-based evaluation
behavior~(\S\ref{sec:exp2-evaluation-shift}).

\section{Experiment 2: Evaluation Shift}
\label{sec:exp2-evaluation-shift}

This experiment addresses RQ2 by testing whether candidate answer-side
intrusion predicts retrieval-based evaluation shift relative to the
observed human-query distribution. We measure pool coverage, judged
ratio, system ranking correlation, and metric gap behavior across
prompt conditions and retrieval systems, then assess the incremental
contribution of concept-level HCIR after controlling for query length
and condition.

\begin{table}[t]
\centering
\footnotesize
\caption{Evaluation shift across prompt conditions (BM25, 100~topics,
77K~LLM + 11K~human queries). Patterns replicate across BM25+RM3, QL, and a cross-encoder
reranker (see text). Judged ratio measures pool coverage;
nDCG@10 and Bpref provide complementary views of effectiveness under
incomplete judgments. Effect sizes (Cohen's $d$) are relative to human
queries. All pairwise differences are significant ($p < 10^{-10}$,
Holm-corrected).}
\label{tab:exp2-evaluation-shift}
\setlength{\tabcolsep}{4pt}
\begin{tabular}{@{}l r r r r r@{}}
\toprule
\textbf{Condition} & \textbf{N} & \textbf{nDCG@10} & \textbf{Bpref} & \textbf{Judged@10} & \textbf{$d$ (judged)} \\
\midrule
human               & 10,835 & 0.203 & 0.169 & 0.914 & --- \\
\midrule
boundary            & 13,819 & 0.180 & 0.192 & 0.738 & $-1.08$ \\
generic             & 15,713 & 0.161 & 0.213 & 0.585 & $-1.74$ \\
high\_knowledge     & 15,990 & 0.140 & 0.214 & 0.473 & $-2.14$ \\
oracle\_text        & 15,697 & 0.178 & 0.238 & 0.574 & $-1.82$ \\
oracle\_terms       & 15,785 & 0.182 & 0.244 & 0.589 & $-1.76$ \\
\bottomrule
\end{tabular}
\end{table}

\textit{Evaluation shift is real and large.}
Table~\ref{tab:exp2-evaluation-shift} reports retrieval behavior on
BM25 for the observed human-query distribution and five LLM prompt
conditions. The judged ratio---the fraction of retrieved documents
covered by existing relevance judgments---drops from 0.91 for human
queries to 0.47 for high-knowledge and 0.59 for generic conditions
(Cohen's $d$~\citep{cohen1992power} ranging from $-1.08$ to $-2.14$, $p < 10^{-10}$ for all
pairwise comparisons). Boundary-constrained prompts reduce the shift
moderately ($d = -1.08$), maintaining a judged ratio of 0.74, but do
not restore human-range behavior. The ranking shift is equally
pronounced: mean rank-biased overlap~\citep{webber2010similarity} (RBO$_{\min}$, $p{=}0.9$) between
human-query rankings and LLM-query rankings ranges from 0.23
(boundary) to 0.10 (high-knowledge), well below the conservative
threshold of 0.5. LLM queries thus retrieve from
substantially different corpus regions, with incomplete
judgment coverage amplifying the divergence.

\textit{\Zthreeauto{} intrusion explains negligible variance.}
A mixed-effects regression (metric\,${\sim}$\,HCIR\,+\,condition\,+\,length\,+\,(1\,|\,topic))
tested the incremental contribution of concept-level HCIR after
controlling for prompt condition and query length. For nDCG@10,
condition alone explains 0.9\% of variance; adding HCIR increases
$R^{2}$ to 2.1\%, yielding an incremental $R^{2}$ of 1.2\%.
HCIR's coefficient is statistically significant (nDCG:
$\beta = +0.064$, $p < 10^{-10}$; judged ratio: $\beta = +0.112$,
$p < 10^{-10}$), but the effect size is practically negligible.
Condition gradient and query length together account for the bulk of
evaluation shift; candidate answer-side intrusion, operationalized at
the concept level, does not meaningfully predict retrieval metric
differences. Under the tested conditions, \Zthreeauto{} density is not a primary driver
of aggregate evaluation distortion.

\textit{Bpref narrows the gap, confirming conservative bias.}
Bpref, designed to tolerate incomplete judgments~\citep{buckley2004bpref}, consistently narrows
the human--LLM gap relative to nDCG@10. Where nDCG@10 drops by
0.063 points from human to high-knowledge queries, bpref \emph{rises}
by 0.046 points, producing a 173\% gap reversal. For boundary queries,
the pattern replicates: nDCG@10 declines by 0.022, while bpref
increases by 0.024 (206\% reversal). Oracle conditions show even
stronger reversals (380--461\%). This directional shift is consistent
with the pool coverage mechanism: LLM queries retrieve
relevant-but-unjudged documents excluded from the original pooling
process. The nDCG@10 decline reflects underestimation, not
over-retrieval of irrelevant material.

\textit{Length does not confound shift.}
The mixed-effects model includes query length as a covariate; the
condition gradient survives this control. Length-stratified
analysis---comparing human and LLM queries matched on token
count---preserves the judged-ratio deficit, confirming that
evaluation shift reflects vocabulary divergence rather than a
specificity artifact.

\textit{Robustness across systems.}
Core patterns replicate across BM25, BM25+RM3, query likelihood, and a
cross-encoder reranker re-scoring BM25 top-100 candidates: condition
gradients in judged ratio and nDCG@10 maintain identical rank order across
all four systems, boundary prompts consistently outperform other LLM
conditions, and HCIR incremental $R^{2}$ remains below 2\% regardless of
scoring function.
Topic-level system-ranking correlation (Kendall's~$\tau$~\citep{kendall1938new} over three
lexical systems per topic) declines from boundary ($\bar{\tau} = 0.49$, 81\%
positive) to high-knowledge ($\bar{\tau} = 0.31$, 70\% positive),
tracking the evaluation-shift gradient. Under lexical systems, bpref
narrows the human--LLM gap as reported above; the cross-encoder reverses
this pattern (human bpref 0.209 vs.\ LLM conditions 0.140--0.187),
because neural re-scoring promotes unjudged documents from BM25's
candidate pool for LLM queries, amplifying rather than smoothing the pool
coverage mismatch.

Having established that evaluation shift is driven by vocabulary
divergence and pool coverage mismatch rather than candidate answer-side
intrusion specifically, we turn to localized diagnostic interventions
to test whether \Zthreeauto{} concepts have detectable retrieval
effects at the individual query level~(\S\ref{sec:exp3-diagnostic-intervention}).

\section{Experiment 3: Diagnostic Intervention}
\label{sec:exp3-diagnostic-intervention}

Experiment 2 established that evaluation shift between human and LLM
queries is real and large but driven by vocabulary divergence and pool
coverage mismatch, with \Zthreeauto{} intrusion explaining negligible
aggregate variance ($R^{2} < 2$\%). However, negligible aggregate variance does not preclude
localized retrieval effects at the individual query level. Here we
employ two diagnostic probes---deletion and injection---to test whether
\Zthreeauto{} concepts carry retrieval signal that distinguishes them
from human-tail concepts or random query components. These are
controlled interventions on synthetic queries, not naturalistic
distribution shifts, designed to isolate \Zthreeauto{} contributions
beyond length, IDF, and rarity.

\textit{Deletion: \Zthreeauto{} concepts carry disproportionate retrieval signal.}
We deleted all \Zthreeauto{} surface forms from 14{,}755 queries
containing at least one \Zthreeauto{} concept, producing three paired
conditions: (i)~\emph{\Zthreeauto{}-deleted}, the modified query; (ii)~\emph{random-deleted},
with the same number of non-\Zthreeauto{} content words removed; and
(iii)~\emph{\Zthreeauto{}-replaced}, with \Zthreeauto{} surface forms replaced by
IDF-matched \Zone{}/\Ztwo{} concepts from the same topic. Paired retrieval on
BM25, BM25+RM3, and query likelihood across ClueWeb12-B13 yields
44{,}262 query--system pairs (14{,}755 queries $\times$ 3 systems).
Deleting \Zthreeauto{} concepts reduces nDCG@10 by 0.066 points
relative to the original queries
(Cohen's $d = -0.47$~\citep{cohen1992power}, $p < 10^{-10}$), with effects of comparable
magnitude for Recall@1K ($d = -0.69$) and Judged@10 ($d = -0.66$). The
effect hierarchy is strictly ordered: \Zthreeauto{}-deleted underperforms
random-deleted by $d = -0.34$ ($p < 10^{-10}$), indicating that
\Zthreeauto{} words carry more retrieval signal than arbitrary query
terms. \Zthreeauto{}-replaced partially recovers performance ($d = -0.21$ relative
to \Zthreeauto{}-deleted, $p < 10^{-10}$), confirming that IDF-matched human-range
substitutes restore some but not all retrieval utility. All six pairwise
comparisons remain significant under Holm correction~\citep{holm1979simple}.

\textit{Injection confirms answer-side specificity.}
We appended the highest-IDF \Zthreeauto{} concept for each topic
into 289 queries initially containing zero
\Zthreeauto{} concepts ($N = 867$ query--system pairs across three
systems). Injection produces a small nDCG@10 decline
($d = -0.12$, $p = 0.005$), with a sharper drop in Judged@10 (0.664
to 0.576, $d = -0.36$, $p < 10^{-8}$). Injecting \Ztwo{} concepts or
IDF-matched \Zone{}/\Ztwo{} produces near-zero nDCG@10 effects ($d \approx 0$),
confirming that the injection penalty is \Zthreeauto{}-specific.
In contrast, bpref \emph{increases} under \Zthreeauto{} injection
($d = +0.11$, $p < 0.001$) while other metrics decline---consistent with
the pool coverage mechanism from Experiment~2: \Zthreeauto{} concepts
steer retrieval toward relevant documents outside the original pooling
boundary.

\textit{Dose-response and HCIR correlation.}
The deletion effect exhibits dose-response structure. Queries with one
\Zthreeauto{} concept drop from nDCG@10 of 0.165 to 0.117
($\Delta = -0.048$); those with $\geq 2$ drop from 0.208 to 0.116
($\Delta = -0.092$)---nearly double the magnitude. Both converge to
${\sim}0.116$ after deletion, suggesting \Zthreeauto{} removal eliminates
a document-specific retrieval advantage. Query-level HCIR correlates
negatively with degradation magnitude (Spearman $\rho = -0.34$~\citep{spearman1904proof}, $p < 0.001$,
$N = 14{,}754$).

\textit{Reconciliation with aggregate evaluation shift.}
Three mechanisms explain why localized effects ($d = -0.47$) wash out to
negligible aggregate variance ($R^{2} < 2$\%): (1)~minority prevalence
(${\sim}32\%$ of non-oracle queries contain \Zthreeauto{}), (2)~bidirectional
effects (deletion and injection hurt different query subsets), and (3)~pool
coverage confounds (the bpref anomaly confirms retrieval of relevant-but-unjudged
documents). The construct value is diagnostic, not predictive of system-level
evaluation distortion.

\begin{figure}[t]
  \centering
  \includegraphics[width=\columnwidth]{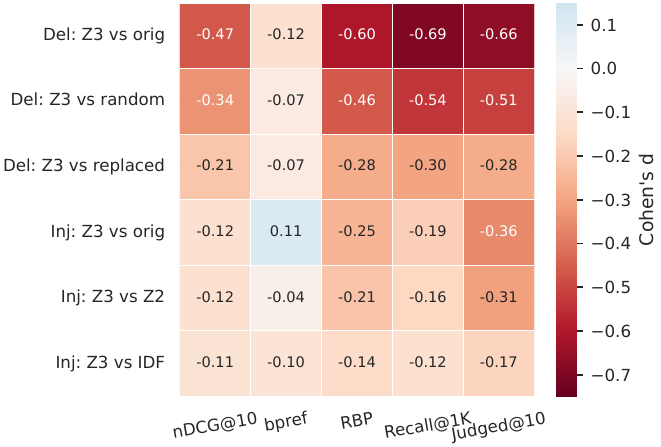}
  \caption{Effect sizes (Cohen's $d$) for deletion and injection probes
    across retrieval and judgment-coverage metrics. Deleting
    \Zthreeauto{} concepts produces consistently larger losses than random
    deletion or IDF-matched replacement, especially for Recall@1K and
    Judged@10. Injection effects are smaller but remain \Zthreeauto{}-specific;
    the positive bpref cell under injection reflects the pool-coverage
    mechanism discussed in Experiment~2.}
  \Description{Heatmap of Cohen's d values for six diagnostic comparisons
    across nDCG@10, bpref, RBP, Recall@1K, and Judged@10. Deletion rows are
    mostly negative, with strongest effects for recall and judged coverage;
    the \Zthreeauto{} injection row has a positive bpref value but negative values for
    other metrics.}
  \label{fig:exp3-metric-effects}
\end{figure}

\textit{\Zthreeval{} directional stability.}
All \Zthreeauto{} items share the document-salience criterion and 0\%
human usage, so deletion and injection effects operate through salience
rather than epistemic judgment. The \ZthreeautoSpecific{} tier
(IDF\,$\geq 3.0$, relaxed precision 78.2\%) provides a high-precision
proxy; the deletion hierarchy (\Zthreeauto{} $>$ random $>$ replacement)
is consistent across all three lexical retrieval systems; cross-encoder
reranking was not applied to diagnostic probes.

Having established that \Zthreeauto{} concepts have medium-to-large
localized retrieval effects that are separable from human-tail concepts
and random query components, we next examine whether these boundary
violations can be mitigated through prompt-based constraints
(\S\ref{sec:exp4-boundary-compliance}).

\section{Experiment 4: Boundary Compliance}
\label{sec:exp4-boundary-compliance}

The preceding experiments established that \Zthreeauto{} concepts
appear at non-trivial rates, resist aggregate-level detection, and
carry localized retrieval signal. We now address RQ3: does prompt-based
mitigation suffice, or is a post-generation validation step required?

\subsection{Prompt-Based Mitigation}
\label{sec:boundary-compliance}

The condition gradient reported in Experiment~1
(Figure~\ref{fig:exp1-condition-gradient}) provides a direct test. Among
non-oracle conditions, boundary-constrained prompts yield the lowest
HCI-Presence (20.7\% vs.\ 33.5\% generic) and a modest concept-HCIR
reduction ($d = -0.04$), confirming that explicit instructions to avoid
unsupported terms have a measurable effect. High-knowledge prompts
achieve a larger HCIR reduction ($d = -0.12$) but increase
HCI-Presence to 41.5\%---a denominator-dilution artifact driven by
longer queries (\S\ref{sec:exp1-findings}), which lower the per-query
\Zthreeauto{} ratio even as more queries contain at least one
candidate answer-side concept. Neither strategy eliminates intrusion:
even under the best-performing boundary condition, one in five queries
contains a \Zthreeauto{} concept.

Cross-model variation reinforces this conclusion: topic accounts for
${\sim}67\%$ of concept-HCIR variance, with condition and model each
below 1\%~(\S\ref{sec:exp1-findings}). Under the tested conditions,
prompt engineering reduces but does not eliminate intrusion.

\subsection{Concept-Provenance-Constrained Selection}
\label{sec:exp6-selection}

Rather than constraining generation, we apply the concept-provenance
framework as a post-generation filter. For each of 100~topics, we
select from 455--770 existing candidates (eight models $\times$ three
non-oracle conditions $\times$ 20~candidates per setting) using a
greedy min-\Zthreeauto{} strategy that minimizes \Zthreeauto{} count,
breaking ties by HCIR.

This strategy achieves \Zthreeauto{}\,=\,0 in 99 of 100~topics under
Pipeline~A, reducing mean HCIR from 6.23\% (random selection) to
0.06\%. Adding a human-central-rate quality floor ($\geq$~median HCR)
preserves the same 99\% zero-intrusion coverage while improving HCR
from 0.468 to 0.612---selecting queries that are
boundary-compliant and better aligned with human-central vocabulary.
Pipeline~B independently achieves 100\% zero-intrusion across all
topics; cross-pipeline agreement holds for 99 of 100~topics.
However, unconstrained min-\Zthreeauto{} selection entails a utility
trade-off: relative to random selection, human utility percentile
drops from 0.438 to 0.382 and nDCG@10 from 0.176 to 0.157 under BM25.
The quality floor recovers most of this loss---utility percentile
reaches 0.417, nDCG@10 reaches 0.174, and judged@10 improves from
0.577 to 0.627---an improved trade-off between
boundary compliance and query utility under the tested candidate pool. Selection efficiency diminishes rapidly with pool size: five candidates
per topic already suffice for 95\% zero-intrusion coverage, and
expanding to 20~candidates improves this to only 97\%.

\begin{figure}[t]
  \centering
  \includegraphics[width=\columnwidth]{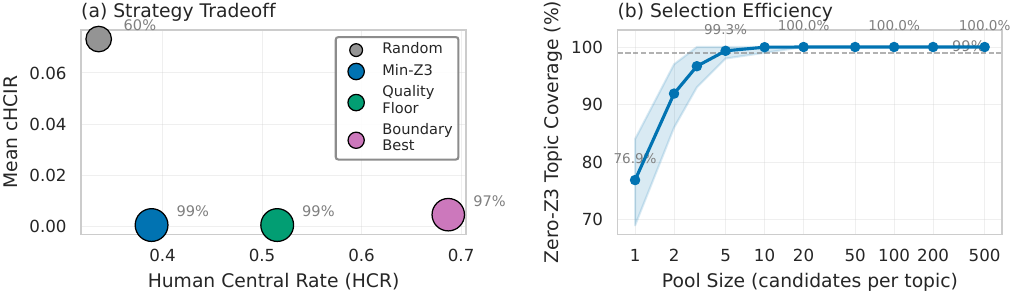}
  \caption{Concept-provenance-constrained selection. (a)~Strategy comparison
    by mean concept-HCIR and HCR; labels show zero-\Zthreeauto{} topic
    coverage. (b)~Coverage rises rapidly with modest over-generation.}
  \Description{Two-panel figure. Left: scatter of HCIR vs HCR for random,
    min-\Zthreeauto{}, quality-floor, and boundary-best strategies. Right: zero-\Zthreeauto{}
    coverage vs candidates per topic.}
  \label{fig:exp6-selection}
\end{figure}

Only one topic (UQV100.052) resists elimination across all 457
Pipeline~A candidates (minimum \Zthreeauto{}\,=\,1).

To close this gap, we re-generated queries for all 40~topics with
\Zthreeauto{}\,$>$\,0 in the original pool, adding explicit negative
constraints (per-topic concept avoidance lists) to the generation
prompt. Across eight models $\times$ three candidates $\times$
40~topics (930~evaluated generations), 94.9\% achieve
\Zthreeauto{}\,=\,0; every topic yields at least one zero-intrusion
candidate, and UQV100.052 is now fully \Zthreeauto{}-free. Combined
with selection, coverage reaches 100\% (40/40~topics). Constraint
compliance varies with model capacity: \texttt{gpt-4.1} and
\texttt{gpt-5.4} reach 98.3\% zero-\Zthreeauto{} rates, whereas
\texttt{gpt-5.4-nano} achieves only 86.7\% (mean
\Zthreeauto{}\,=\,0.583). The most resistant topics (UQV100.074,
UQV100.016) both arise under high-knowledge conditions, suggesting
that complex information needs are harder to constrain through negative
prompting.

Over-generation followed by
concept-provenance-constrained selection---augmented by targeted
re-generation for resistant topics---nearly eliminates candidate
answer-side intrusion. Combined
with a human-central-rate quality floor, this strategy recovers most utility losses,
a trade-off that prompt engineering alone does not achieve under the tested conditions.

Having shown that concept-provenance analysis both diagnoses boundary
violations and enables their near-elimination through constrained
selection, we discuss the broader implications for synthetic query
validation (\S\ref{sec:discussion}).

\section{Discussion}
\label{sec:discussion}

Concept provenance reveals a mismatch between local and aggregate
effects. Deleting \Zthreeauto{} concepts produces a medium localized
retrieval effect (nDCG@10 $d = -0.47$~\citep{cohen1992power}), exceeding
random-word deletion ($d = -0.34$) and IDF-matched replacement
($d = -0.21$). Yet their density explains less than 2\% of incremental
aggregate metric variance after controlling for condition and query length.
Its value is therefore diagnostic: it identifies \emph{which} queries violate the
information-access boundary rather than predicting aggregate evaluation
shift. Constrained selection (\S\ref{sec:exp6-selection}) makes this
diagnosis actionable, achieving 99\% elimination across 100~topics, but
does not by itself solve query simulation.

\textit{Dual intrusion mechanism.}
Human validation decomposes \Zthreeauto{} into two distinct
mechanisms. Of 220~validated items, 45.5\% were classified as
\emph{requires\_research}---concepts genuinely unknown to a general
pre-search user. This \emph{knowledge intrusion} predominantly involves
answer-side entity types: named entities, technical vocabulary, and
mechanism terms constitute 73.7\% of confirmed cases, consistent with
LLMs drawing on parametric knowledge of document content. A further
45.0\% were \emph{common\_knowledge} or \emph{personal\_experience}:
concepts a user could know, but would more plausibly discover through
browsing than deploy in an initial query. This \emph{deployment
intrusion} is behavioral rather than epistemic, but can likewise steer
retrieval toward answer-relevant documents. The remaining 9.5\% are
backstory false negatives, predominantly compositional cases such as
``fistula treatment'' when ``fistula'' and ``treatment'' occur separately;
BCC reclassifies these cases (\S\ref{sec:exp1-robustness}). The near-equal
split between knowledge and deployment intrusion shows that \Zthreeauto{}
captures boundary violations more broadly than unfamiliar vocabulary
alone. Strict precision (45.5\%) measures knowledge intrusion, whereas
relaxed precision (68.2\%) captures both mechanisms. Both can distort
evaluation; the positive bpref response to injection ($d = +0.11$) is
consistent with retrieving relevant but unjudged documents.

\textit{LLM knowledge projection bias.}
Human validation also reveals that LLMs systematically
conflate ``I know this concept'' with ``most people know this
concept.'' When classifying the same concept--query pairs annotated by
humans, six LLM annotators downgraded 16--32\% of human-confirmed
\emph{requires-research} items to \emph{common knowledge}. LLM--LLM
agreement was high (Cohen's $\kappa$~\citep{cohen1960coefficient} up to 0.791), but LLM--human
agreement was only fair ($\kappa \approx 0.36$), indicating systematic
calibration error rather than random noise. This knowledge projection
bias means that LLM annotation cannot substitute for human judgment in
concept-origin classification.

\textit{Operational implications.}
Concept provenance complements diversity, effectiveness, and utility
metrics~\citep{alaofi2023,zendel2025,ran2025,breuer2024,kruff2026} by adding alignment with the simulated searcher's
information-access boundary. Constrained selection
(\S\ref{sec:exp6-selection}) demonstrates a practical deployment pathway:
over-generate, filter by provenance, and select boundary-compliant
variants. A human-central-rate quality floor recovers most utility
losses from bare minimization~(\S\ref{sec:exp6-selection}), but the
threshold was fixed at pool median and not sensitivity-tested; whether
alternative floors or retrieval configurations shift the trade-off
frontier remains open. Selection also changes the ecological query
distribution by privileging candidates that pass a provenance filter; we
therefore view it as a compliance mechanism for evaluation construction,
not as evidence that the selected queries fully reproduce the natural human
query distribution.

\section{Limitations and Conclusion}
\label{sec:limitations-conclusion}

Several methodological boundaries and scope constraints qualify the findings.

\textit{Construct validity.}
Provenance zones are relative to the observed UQV100 human-query
distribution (49--250 variants per topic, median~77), not to
population-complete ground truth. Additional human sampling could move some
\Zthreeauto{} concepts into \Ztwo{}, so our claims concern LLM
\Zthreeauto{} rates, their variation, and their retrieval effects under the
tested conditions, not human-versus-LLM superiority. Backstory matching
requires explicit textual presence; human validation identified 9.5\% of
\Zthreeauto{} items as backstory false negatives, most matching the BCC
pattern. BCC-aware matching reduces \Zthreeauto{} by 10.7\% and mean HCIR
by 15\%~(\S\ref{sec:exp1-robustness}), confirming compositional coverage
as a tractable refinement; the primary operationalization retains strict
matching for \Zzero{} precision. \Zthreeauto{} strict precision (45.5\%) falls
below the pre-registered 50\% threshold, placing the automatic label in
the weakened tier. We therefore treat it as a high-recall diagnostic,
with relaxed precision (68.2\%) capturing both intrusion mechanisms; the
high-specificity subset reaches 78.2\%. Condition gradients and intervention
directions remain stable on validated items because all \Zthreeauto{} concepts
share the document-salience property through which the interventions operate.
Concept-level HCIR is a conservative lower bound because it misses compositional violations.
General-population calibration also ignores implied searcher expertise and
may over-count intrusions for expert personas.

\textit{Data and generalizability.}
The evaluation relies on a single collection (UQV100~\citep{bailey2016}: 100 topics,
crowdworker queries, ClueWeb12-B13). Operationalizing the \Zone{}/\Ztwo{}/\Zthreeauto{} distinction requires
sufficient independent human query variants per topic. UQV100 provides
${\sim}57$ unique variants per topic across 263 workers---to our
knowledge, the only publicly available IR test collection at this
per-topic density. Generalizability is addressed through internal
robustness: 8 models across 4 families, 5 conditions including oracle
controls, three lexical retrieval systems plus a cross-encoder reranker for
ranker-type robustness, dual extraction pipelines, and threshold sensitivity across five
pre-registered parameters. Incomplete relevance
judgments~\citep{buckley2004bpref} bias \Zthreeauto{} toward
under-detection; TF-IDF cutoff sensitivity ($k \in \{10\text{--}200\}$)
confirms stable condition gradients and a median of 98.5 relevant
documents per topic mitigates sparsity. The reranker checks ranker-type robustness for evaluation shift and
selection (Experiments~2 and~4), but oracle conditions and diagnostic
probes (Experiment~3) were not evaluated under neural re-scoring; it is
not a standalone dense retrieval experiment. Constructing a second collection at
this per-topic density is itself a substantial effort and a priority for
future work.

\textit{Method.}
Both pipelines share TF-IDF-based salience and may agree on false
positives; token-level HCIR convergence ($\rho = 1.0$) and manual
annotation mitigate but cannot rule out shared errors. The observed
knowledge-projection bias (LLM--LLM $\kappa$ up to 0.791 vs.\ approximately
0.36 LLM--human) also precludes substituting LLM annotators for human judgment.

\textit{Scope.}
Eight models across four LLM families were tested, but findings may not
generalize to all current or future architectures. The framework addresses
initial query formulation only ($B_t \to q$); it does not extend to query
reformulation or session-level
behavior~\citep{engelmann2024context,ren2024bases,zhang2024usimagent}.
Training contamination cannot be
ruled out, but our framing treats this as operationally irrelevant: the
evaluation-validity consequence is the same whether intrusion arises from
memorization or instruction-following. Deletion and injection probes are
artificial interventions rather than naturalistic queries; we use hedged
language throughout to avoid causal overclaims. The injection probe uses the
highest-IDF \Zthreeauto{} concept per topic as a stress test; sampling
multiple or median-salience concepts would better characterize the
intervention distribution.

\textit{Conclusion.}
This paper establishes that LLM-generated initial search queries can
violate the simulated searcher's information-access boundary through
candidate answer-side concept intrusion---a phenomenon that existing
query-generation metrics~\citep{alaofi2023,zendel2025,kruff2026} cannot
detect. Concept provenance separates human-tail variation from knowledge
and deployment intrusion. The experiments show that such concepts have
localized retrieval effects but little aggregate predictive power, making
provenance a boundary-compliance diagnostic rather than an
evaluation-shift predictor. Prompting reduces but does not eliminate the
problem; under the tested candidate pool, constrained selection nearly
eliminates detected intrusion while a quality floor recovers most lost utility.
Extending this validation to additional
collections with sufficient per-topic human-variant density and to
session-level simulation remains important future work.


\begin{acks}
This work was conducted at the ARC Centre of Excellence for Automated Decision-Making and Society (ADM+S) and funded by the Australian Research Council (CE200100005). It was supported by computing resources from RACE (RMIT Advanced Computing Ecosystem). 
\end{acks}

\newpage
\section*{GenAI Usage Disclosure}

This research complies with the CIKM 2026 GenAI usage policy. The authors disclose the following use of Generative AI (GenAI) tools during the research process:

\begin{itemize}
  \item \textbf{Writing:} ChatGPT (OpenAI) was used to assist in proofreading, rephrasing technical sentences, and improving the clarity of the manuscript. All substantive content, including ideas, methods, results, and analysis, was written and verified by the authors.
  \item \textbf{Code:} No GenAI tools were used to generate any code in this research. All code was developed by the authors.
  \item \textbf{Data:} No GenAI tools were used to generate or augment the data used in this research. All datasets were obtained from publicly available sources as described in the paper.
  \item \textbf{Experiments and Analysis:} No GenAI tools were used to generate experimental results or statistical analyses.
\end{itemize}

The authors confirm that all intellectual contributions are original and that the use of GenAI tools did not compromise the scientific integrity or originality of the work.

\bibliographystyle{ACM-Reference-Format}
\balance
\bibliography{refs}

\end{document}